\documentclass[10pt]{article}
\usepackage{fa2023}
\usepackage{amsmath}
\usepackage{cite}
\usepackage{url}
\usepackage{graphicx}
\usepackage{color}
\usepackage{siunitx}
\usepackage[utf8]{inputenc}
\usepackage{xcolor}
\usepackage{multirow}
\usepackage{subfigure}
\renewcommand*{\vec}[1]{\bm{\mathrm{#1}}}

\title{Influence of Lossy Speech Codecs on Hearing-aid, Binaural Sound Source Localisation using DNNs}

\oneauthor
{Siyuan Song \ \  Stijn Kindt \ \ Jasper Maes \ \ Alexander Bohlender \ \ Nilesh Madhu}{
  IDLab, Department of Electronics and Information Systems, Ghent University - imec, Belgium\\
 \{firstname\}.\{lastname\}@ugent.be, jantmaes.maes@ugent.be\\
\correspondingauthor{siyuan.song@ugent.be}{Siyuan Song et al.}
}

\begin{document}

\maketitle
\begin{abstract}
Hearing aids are typically equipped with multiple microphones to exploit spatial
information for source localisation and speech enhancement. Especially for hearing
aids, a good source localisation is important: it not only guides source
separation methods but can also be used to enhance spatial cues, increasing user-awareness of important events in their surroundings. We use a state-of-the-art deep neural network (DNN) to perform binaural direction-of-arrival (DoA) estimation, where the DNN uses information from all microphones
at both ears. However, hearing aids have limited bandwidth to exchange this data.
Bluetooth low-energy (BLE) is emerging as an attractive option to facilitate such
data exchange, with the LC3plus codec offering several bitrate and latency trade-off possibilities. In this paper, we investigate the effect of such lossy codecs
on localisation accuracy. Specifically, we consider two conditions: processing at one
ear vs processing at a central point, which influences the number of channels that
need to be encoded. Performance is benchmarked against a baseline that allows full
audio-exchange - yielding valuable insights into the usage of DNNs under lossy
encoding. We also extend the Pyroomacoustics library to include hearing-device and
head-related transfer functions (HD-HRTFs) to suitably train the networks. This
can also benefit other researchers in the field.
\end{abstract}
\section{Introduction}\label{sec:introduction}

Sound source localisation plays a vital part in the auditory experience. One example can be in group conversations, when the conversation switches from one speaker to another, the listener needs to locate the new speaker instantly, as otherwise, understanding may be reduced seriously \cite{hawley2004benefit, bronkhorst1988effect}. Especially for a hearing-impaired person, localisation plays a more significant role in speech communication difficulties than is usually appreciated \cite{akeroyd2014overview, 10.1121/1.424672}. Thus algorithms for binaural enhancement of speech in hearing aids are faced with the challenging task of maintaining the spatial cues of target {\em and} interfering sound sources, which requires a good, implicit or explicit, DoA estimation~\cite{van2007binaural}.

For binaural hearing aid-based localisation, one way of estimating the DoAs is by matching the estimated relative transfer functions (RTFs) of the microphones with ideal, anechoic RTFs from each direction~\cite{7336917}. It has been shown that this method outperforms the dual delay line approach \cite{4445829} and pressure-energy gradient approach \cite{ahonen2012parametric}. However, prior knowledge of speaker-specific anechoic head-related transfer function (HRTF) or RTFs from all angles is required for this approach. In \cite{7418338}, another method with the maximum likelihood framework is proposed assuming accessibility of the noise-free version of the target signal. Whereas model-based methods like these strongly rely on prior information and simplified assumptions, data-driven deep learning methods have recently been widely investigated in DoA estimation~\cite{grumiaux2022survey}. In \cite{9381644}, long short-term memory (LSTM) or temporal convolutional network (TCN) are integrated into the convolutional neural network (CNN)-based DoA estimator of \cite{8651493} to exploit the temporal context. This approach was shown to outperform existing state-of-the-art methods including steered response power with phase transform (SPR-PHAT)\cite{DiBiase2000AHL}, informed phase unwrapping (IPU)-least-squares (LS) method\cite{9053359}, and the CNN baseline \cite{DBLP:journals/corr/ChakrabartyH17}. Here, we first extend the CNN/LSTM model of \cite{9381644} for binaural DoA estimation in hearing aids. Since we no longer deal with free-field arrayes, this requires the incorporation of the 
measured, hearing-device and head-related transfer functions (HD-HRTFs) \cite{denk2018adapting} of the respective microphones in the training paradigm. Based on this extension, we investigate the influence of lossy codecs on binaural DoA estimation and evaluate mitigating strategies as well. This forms the key contribution of this work.

Data exchange in hearing aids is often achieved by wireless communication platforms including Bluetooth and Digital Enhanced Cordless Telecommunications. LC3 and its extension LC3plus\cite{schnell2021lc3}, which stands for low complexity communications codec, are promising technologies that aim to provide the solution to transmit high-quality audio over wireless accessories at reduced bandwidth/bitrates\footnote{\url{https://www.iis.fraunhofer.de/en/ff/amm/communication/lc3.html}}. In the application of hearing aids, to get binaural information, the recorded signals from the microphones on the devices can be pooled in two ways: (i) when the device at one ear is responsible for the processing, the signals from the device on the other ear are transmitted to this device, or (ii) when the processing has to be done on an external, central processor, signals from both devices need to be transmitted to this processor. In both cases, a certain amount of audio data needs to go through the codec. As the codec is typically {\em lossy}, this will change the interaural time differences (ITDs) and interaural level differences (ILDs), which are important DoA cues~\cite{blauert1997spatial}, leading to a less accurate localisation -- especially in the case of data-driven approaches. In this paper, both cases are investigated under the use of the lossy BLE LC3plus codec. Next, model training is performed {\em with} the codec in the loop to examine if the degradations caused by the codec can be recovered through (re-)training or if the DoA information is totally lost.

The paper is structured as follows: Section \ref{sec:cnn_model} presents the signal feature and the baseline multi-source binaural DoA estimation based on our state-of-the-art CRNN model.
In Section \ref{sec:codec}, DoA estimation using the two data-exchange paradigms is briefly presented. The experimental setup including microphone array and training data generation is discussed in Section \ref{sec:setup}. In Section \ref{sec:discussion}, first, the performance of the baseline DoA estimator is evaluated under three conditions: (i) when no codec is used (full-bandwidth audio exchange), (ii) when only 3 channels are encoded, and (iii) when all 6 channels are encoded. Next, the performance improvement with codec-in-training-loop is evaluated for the latter two conditions. We conclude with some thoughts for further generalisation and directions for future work.
\section{Multi-source Binaural DOA Estimation}\label{sec:cnn_model}

\subsection{Behind-the-ear Microphone Array}\label{subsec:body}
\begin{figure}[h]
 \centerline{
 \includegraphics[width=0.99\columnwidth]{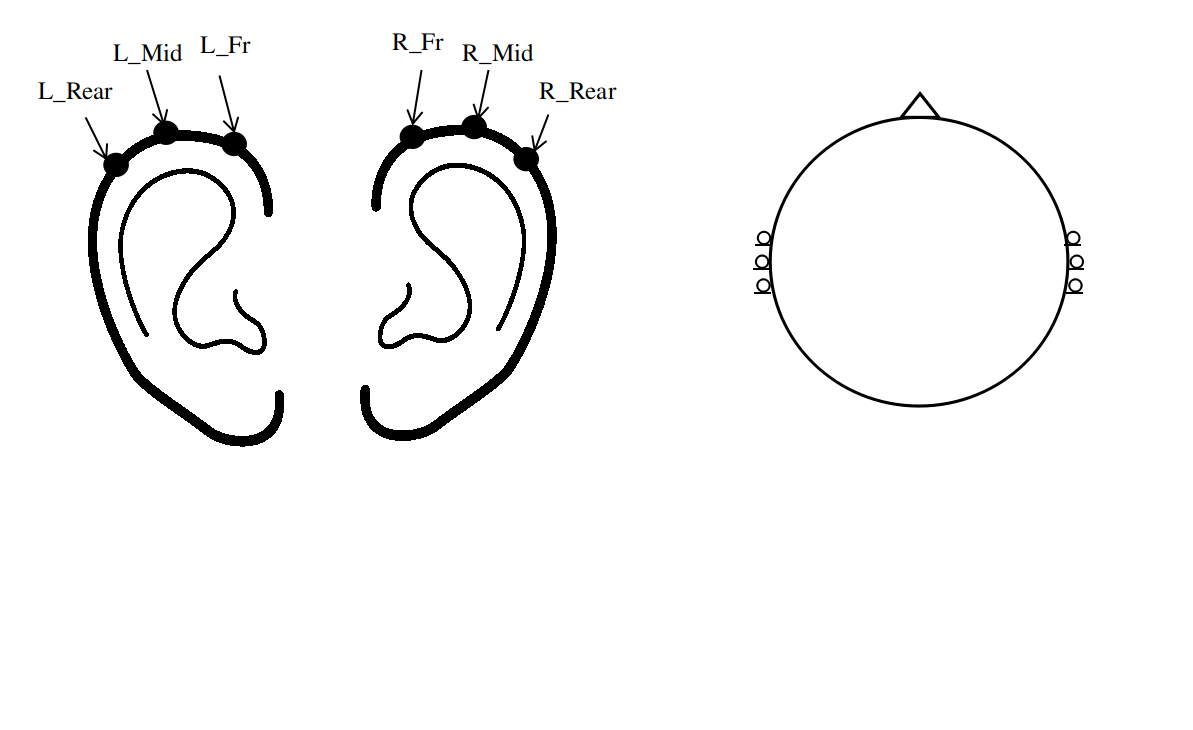}}
\vspace*{-4em}
 \caption{The positioning of two BTE arrays at each ear and channel index of each mirophone}\label{fig:BTE}

\end{figure}

The binaural microphone array adopted in this paper consists of 6 microphones, 3 on each of the two behind-the-ear (BTE) devices~\cite{BTE}. The microphone geometry for each device is shown in Fig.\ref{fig:BTE} where the 3 microphone channels are denoted as: front (\_Fr), middle (\_Mid) and rear (\_Rear). The distance between neighboring microphones on each device is approximately 7.6 mm. This configuration is used for training and testing.

\subsection{Feature extraction for data-driven approaches}

We consider the $N$-channel microphone array signals in the short-time Fourier transformation (STFT) domain, where the signal observed at the $n$-th microphone can be modelled as
\begin{equation}
   Y_{n}(k,m) = \sum_j S_{j, n}(k,m)+V_{n}(k,m).
\end{equation}
${k \in \lbrace1,..., K\rbrace}$ and ${m \in \lbrace1,..., M\rbrace}$ represent the frequency index and frame index respectively. $S_{j,n}(k,m)$ is the STFT representation of the signal from source $j$ at microphone $n$. $V_n(k,m)$ is the additive noise.

In \cite{grumiaux2022survey}, a comprehensive summary of input features  used in source localisation is provided. Consistent with \cite{DBLP:journals/corr/ChakrabartyH17,9381644}, we use the phase component of the magnitude-phase representation of \ref{eq:mag-ph} as input features: 
\begin{equation}
    Y_n(k,m) = |Y_n(k,m)|e^{j \angle Y_n(k,m)}.
    \label{eq:mag-ph}
\end{equation}
To avoid the $2\pi$ phase wrapping problem, we take the sine and cosine of the phase. Additionally, we include the normalized magnitude, giving us the following vector of $3$ elements as input for the $n$th channel:
\begin{equation}
\begin{split}
    \vec{F}_n(k,m) = 
    \left[
    \begin{aligned}
          \sin ( & \angle Y_n (k,m)) \\
          \cos ( & \angle Y_n (k,m)) \\
          |& \overline{Y}_n (k,m)|\\
    \end{aligned}
    \right],
\end{split}
\end{equation}
By normalizing the magnitude with
\begin{equation}
    |\overline{Y}_n(k,m)| = \left.|Y_n(k,m)|\middle/ \left(\frac{1}{N} \sum_{\nu=1}^{N} |Y_{\nu}(k,m)| \right)\right.,
\end{equation}
the \textit{differences between the channels} are captured. 
This is motivated by the important role that the ILDs play in binaural localisation. 

From the feature vectors $\vec{F}_n(k,m)$ for all channels and all frequency bins, we form the tensor $\Psi(m)$ of size ${3 \times N \times K'}$ that serves as input to the DNN, where ${K' = K/2+1}$.

\subsection{CNN-based model architecture}

The azimuth angle range of $0^{\circ}$ to $360^{\circ}$ is divided into $I$ sectors, with a fixed resolution. DoA estimation may then be seen as a {\em classification} problem, where probabilities of source activity are calculated for each sector. In this work, we consider $I=72$, with a resolution of $5^{\circ}$, which can be represented by $\varphi_i$, and $i = 0, 1, ..., 72$.

\begin{figure}[ht]
 \centerline{
 \includegraphics[width=0.8\columnwidth]{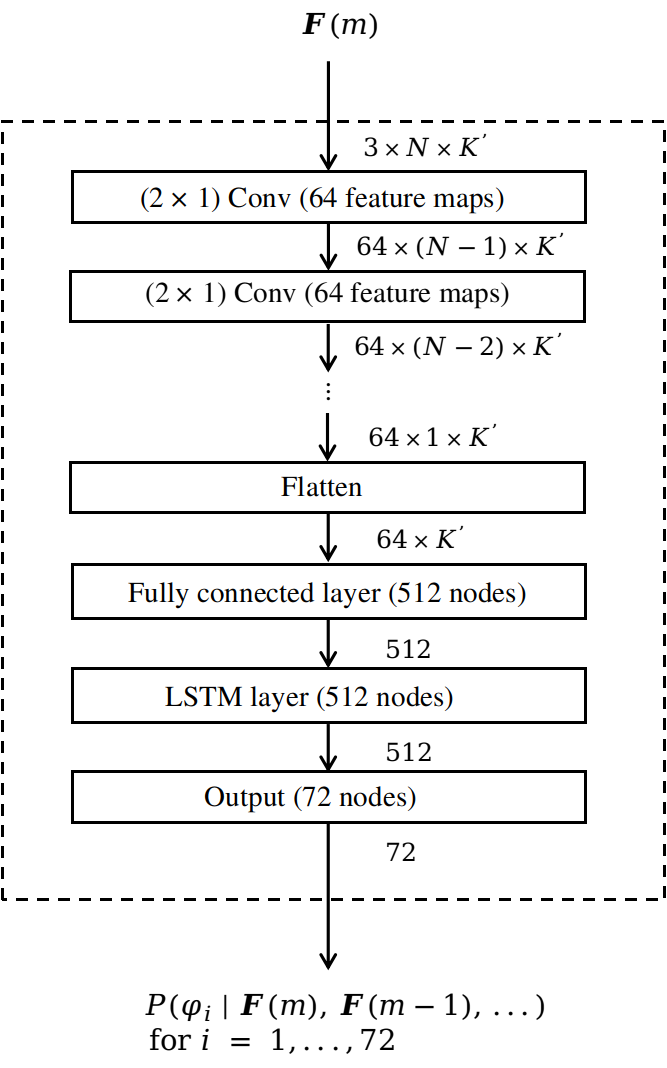}}
 \caption{
 CRNN for binaural sound source localisation.}
 \label{fig:BTE_model}

\end{figure}

The baseline DoA estimation model illustrated in Fig.\ref{fig:BTE_model} is a straightforward extension of the CNN/LSTM model in \cite{9381644} to the binaural case, which is referred to as convolutional recurrent neural network (CRNN) in this paper. The latent feature extraction is done by $(N-1)$ convolutional layers, applied across the channel dimension, separately for each time-frequency bin. One fully connected layer aggregates information along the frequency dimension while the LSTM layer is used to provide temporal context to the network. Note that this model structure is real-time capable since only the STFT operation requires some latency. The final probability of each DoA class is then calculated by the output layer with sigmoid activation. The desired output $P(\varphi_i | F(m), F(m-1), ...) $ is a multi-hot vector consisting of 0 (inactive) and 1 (active) for each DoA.

The binary cross-entropy (BCE) loss is used as the loss function to optimize the weights. For training the model, an AdamW optimizer is used with learning rate 0.0001.  A batch contains 20 signals of 200 frames each.

\section{Influence of LC3plus codec on Localisation Performance}\label{sec:codec}

\begin{figure}[ht]
 \centerline{
 \includegraphics[width=0.6\columnwidth]{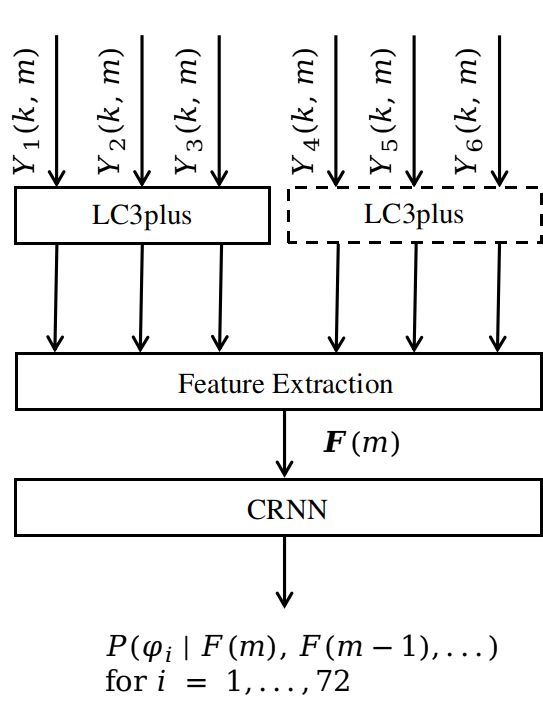}}
 \caption{
 Different ways of encoding: encoding of 3 channels (assuming ecncoding of the right ear) and encoding of 6 channels.
}
 \label{fig:codec}
\end{figure}

Two practical applications are considered for LC3plus codec: one is transferring the observed signals from the device at one ear to the other, and the other is transferring the observed signals from the devices at both ears to a central processor -- as depicted in Fig.\ref{fig:codec}. In both situations, the DoA information contained in the microphone signals is to some extent changed. To evaluate the influence of LC3plus codec, the CRNN with full-bandwidth (unencoded) data exchange is evaluated as the baseline model, under different conditions. Note that the bitrate of 32 kbps are utilized for LC3plus codec (unless stated otherwise).

\vspace{1ex}

\section{Experimental Setup}\label{sec:setup}

\begin{table*}[h]

\caption{Data generation: training, validation and test set}

\begin{center}
\begin{tabular}{lll}

\\ \hline
    \multicolumn{1}{l|}{\multirow{10}{*}{Training set}} &
  \multicolumn{1}{l|}{Speech signals} &
  7438 utterances from TIMIT and PTDB-TUG  dataset \\
\multicolumn{1}{l|}{} & \multicolumn{1}{l|}{Room dimensions (m)}          &  
  R1: \((6 \times 6 \times 2.5)\), 
  R2: \((5 \times 4 \times 2.8)\), 
  R3: \((10 \times 6 \times 2.4)\), 
  \\
\multicolumn{1}{l|}{} & \multicolumn{1}{l|}{}                       &
  R4: \((8 \times 3 \times 3.1)\),
  R5: \((8 \times 5 \times 2.9)\), 
  R6: \((4 \times 9 \times 3.3)\), 
      \\
\multicolumn{1}{l|}{} & \multicolumn{1}{l|}{}          &  
  R7: \((7 \times 7 \times 2.3)\), 
  R8: \((5 \times 6 \times 3.6)\),
  R9: \((9 \times 6 \times 3.2)\), 
  \\
\multicolumn{1}{l|}{} & \multicolumn{1}{l|}{}          &  
  R10: \((11 \times 7 \times 3)\)
  \\
\multicolumn{1}{l|}{} & \multicolumn{1}{l|}{T60}          & 
  R1: \(0.3\) s, R2: \(0.2\) s, R3: \(0.8\) s, R4: \(0.4\) s, R5: \(0.6\) s, R6: \(0.5\) s, 
  \\
 \multicolumn{1}{l|}{} & \multicolumn{1}{l|}{}          & 
  R7: \(0.7\) s, R8: \(0.45\) s, R9: \(0.55\) s, R10: \(0.75\) s
  \\
\multicolumn{1}{l|}{} & \multicolumn{1}{l|}{Array positions}        &   
  7 different positions in each of the rooms
\\
\multicolumn{1}{l|}{} & \multicolumn{1}{l|}{Source-array distances} &      
  \(20\%\), \(40\%\), \(60\%\) and \(80\%\) of the distance between array and wall
\\
\multicolumn{1}{l|}{} & \multicolumn{1}{l|}{Additive noise}         & Simulated diffuse noises with SNRs from 0 to 30 dB \\ \hline
\multicolumn{1}{l|}{\multirow{6}{*}{Validation set}} &
  \multicolumn{1}{l|}{Speech signals} &
  2280 utterances TIMIT and PTDB-TUG  dataset \\
\multicolumn{1}{l|}{} & \multicolumn{1}{l|}{Room dimensions(m)}          & 
    R11: \((5.5 \times 7.5 \times 2.7)\) , 
    R12: \((8.5 \times 4.5 \times 3.5))\),
\\    
\multicolumn{1}{l|}{} & \multicolumn{1}{l|}{}          &  

    R13: \((6.5 \times 6.5 \times 2.3)\) 
\\
\multicolumn{1}{l|}{} & \multicolumn{1}{l|}{T60}          & 
  R11: \(0.525\) s, R12: \(0.625\) s, R13: \(0.475\) s
  \\
\multicolumn{1}{l|}{} & \multicolumn{1}{l|}{Array positions}        &     
4 different positions in each of the rooms \\
\multicolumn{1}{l|}{} & \multicolumn{1}{l|}{Source-array distances} &        
\(30\%\), \(50\%\), \(70\%\) and \(85\%\) of the distance between array and wall\\
\multicolumn{1}{l|}{} & \multicolumn{1}{l|}{Additive noise}         & Simulated diffuse noises with SNRs from 0 to 30 dB \\ \hline
\multicolumn{1}{l|}{\multirow{5}{*}{Test set}} &
  \multicolumn{1}{l|}{Speech signals} &
  1444 utterances from TSP speech database \\
\multicolumn{1}{l|}{} & \multicolumn{1}{l|}{Room dimensions \& T60}          &     
    R14: \((5 \times 4 \times 2.5)\) m,  \(0.2\) s; R15: \((7 \times 6 \times 3.5)\), \(0.6\) s                                          
  \\
\multicolumn{1}{l|}{} & \multicolumn{1}{l|}{Array positions}        &                      
  4 different positions in each room\\
\multicolumn{1}{l|}{} &
  \multicolumn{1}{l|}{Source-array distances} &
  1m, 2m, 3m (Only with the angles inside the room) \\
\multicolumn{1}{l|}{} &
  \multicolumn{1}{l|}{Additive noise} &
  Simulated diffuse noises with SNRs of  {[}5, 10 ,15{]} dB or no noise\\ \hline                                               
\end{tabular}
\end{center}
\label{tab:data_generation}
\end{table*}
All datasets are generated by convolving BRIRs and dry speech signals, with diffuse noise added on top. The sampling rate is fixed at 16 kHz.

\begin{figure*}[h]
\hfill
\subfigure{\includegraphics[width=.405\linewidth]{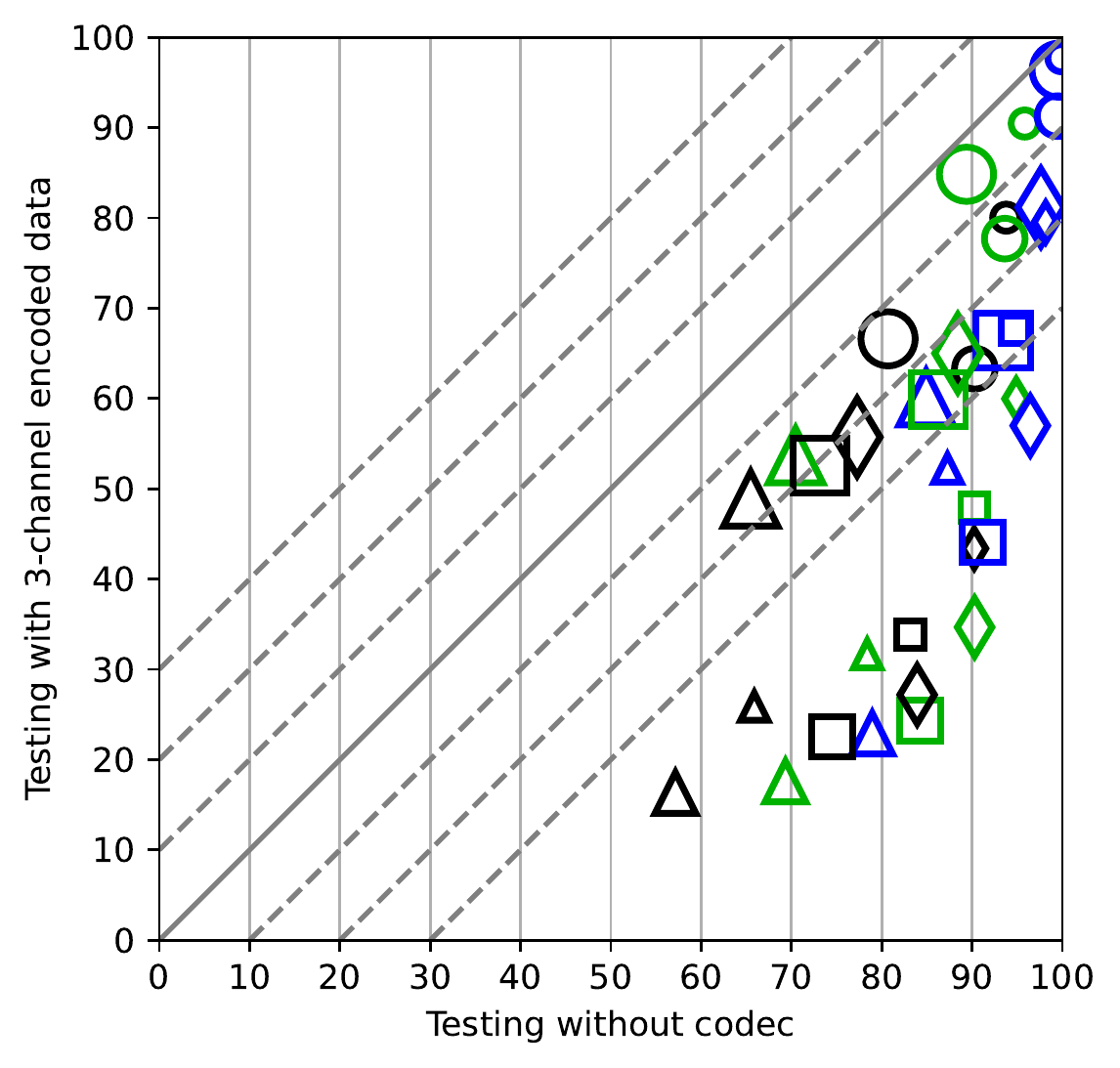}
\label{fig:CNN/LSTM(1)}}
\hfill
\subfigure{\includegraphics[width=.54\linewidth]{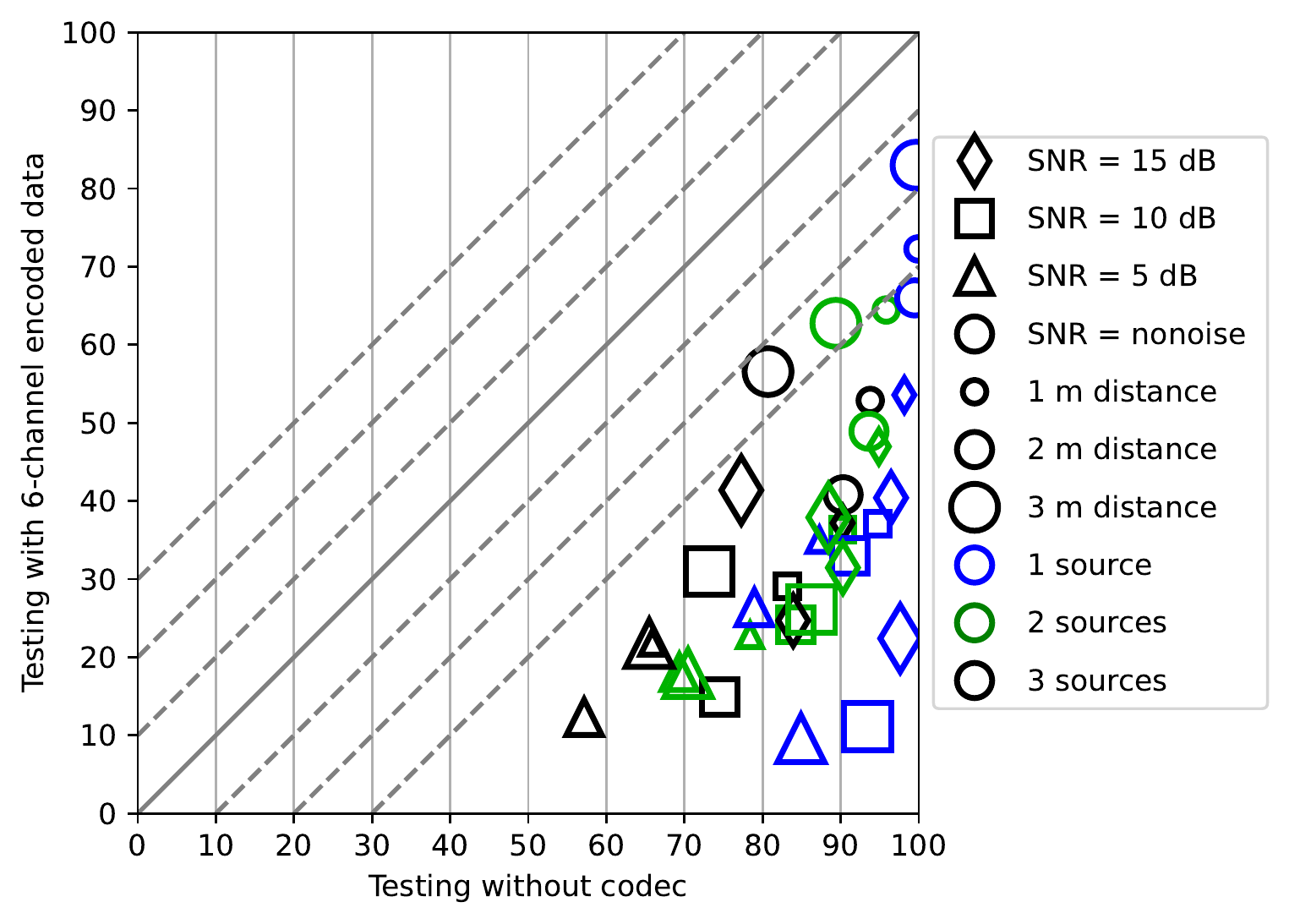}
\label{fig:CNN/LSTM(2)}}

\caption{
\label{fig:CNN/LSTM}
The localisation \textbf{accuracies(\%) }of CRNN baseline in different conditions are indicated in the figure. The x-axis in both subplots is the CRNN baseline tested with unencoded test set. The y-axis in left subplot is the CRNN baseline tested with 3-channel encoded data and the y-axis in the right subplot is tested with 6-channel encoded data.}
\end{figure*}

\begin{figure*}[h]
\hfill
\subfigure{\includegraphics[width=.405\linewidth]{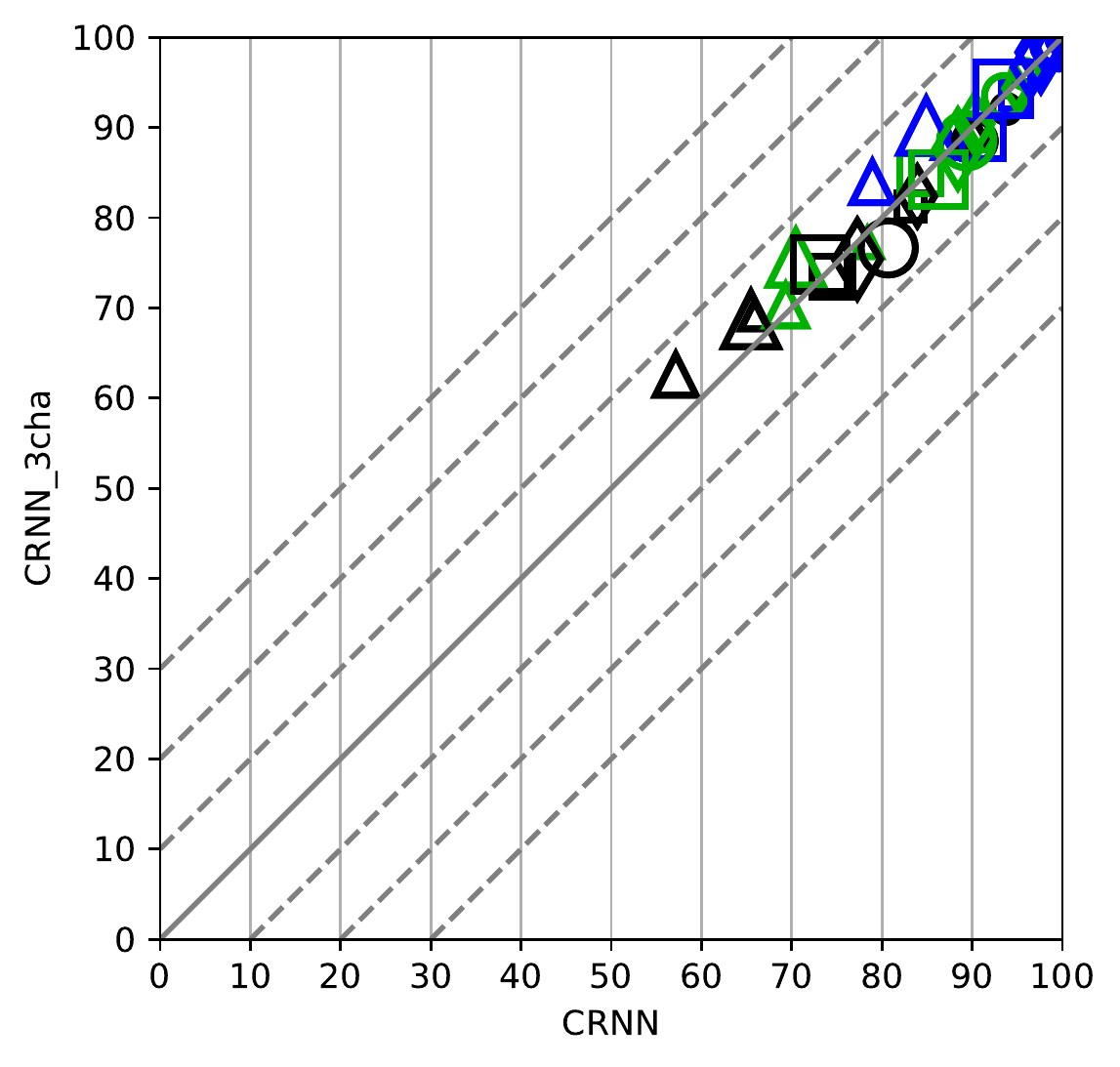}
\label{fig:codec(1)}}
\hfill
\subfigure{\includegraphics[width=.54\linewidth]{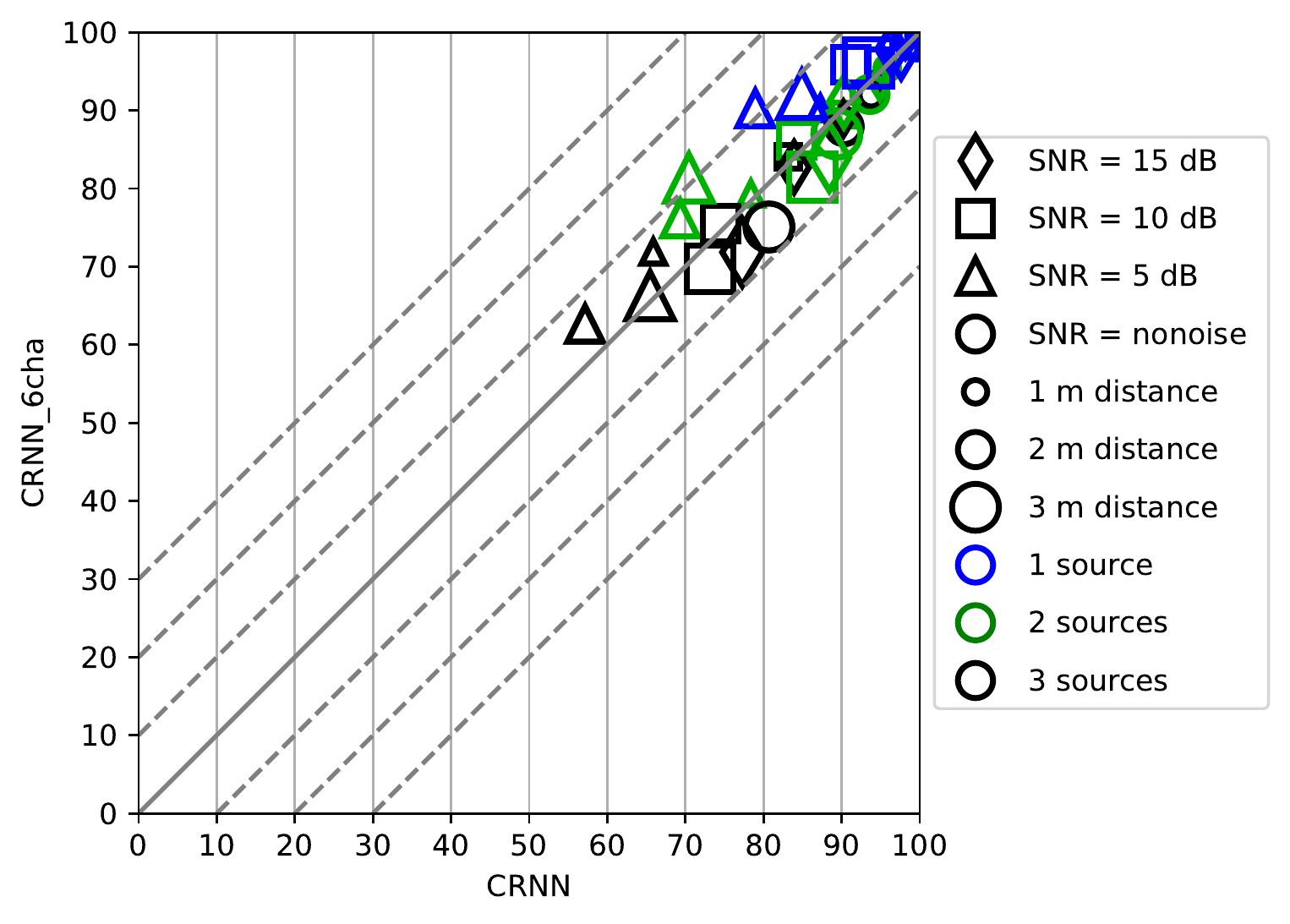}
\label{fig:codec(2)}}

\caption{
\label{fig:codec_results}
The localisation \textbf{accuracies(\%) } of CRNN\_3cha and CRNN\_6cha are indicated in the figure. The x-axis in both subplots is CRNN baseline. The y-axis in the left subplot is CRNN\_3cha and the y-axis in the right subplot is CRNN\_6cha. Note that all the models are tested with corresponding testing conditions (see in section \ref{sec:codec_discussion}).}
\end{figure*}

\begin{figure*}[ht]

\begin{center}
    \subfigure{\includegraphics[width=0.5\linewidth]{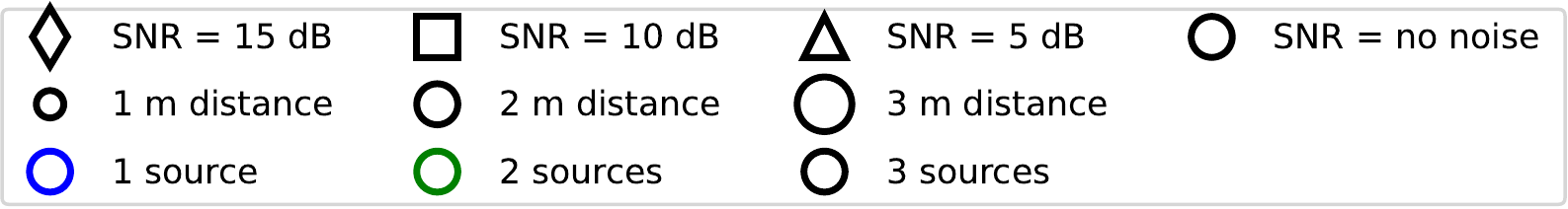}
    \label{fig:legend}}
\end{center}
\subfigure{\includegraphics[width=.33\linewidth]{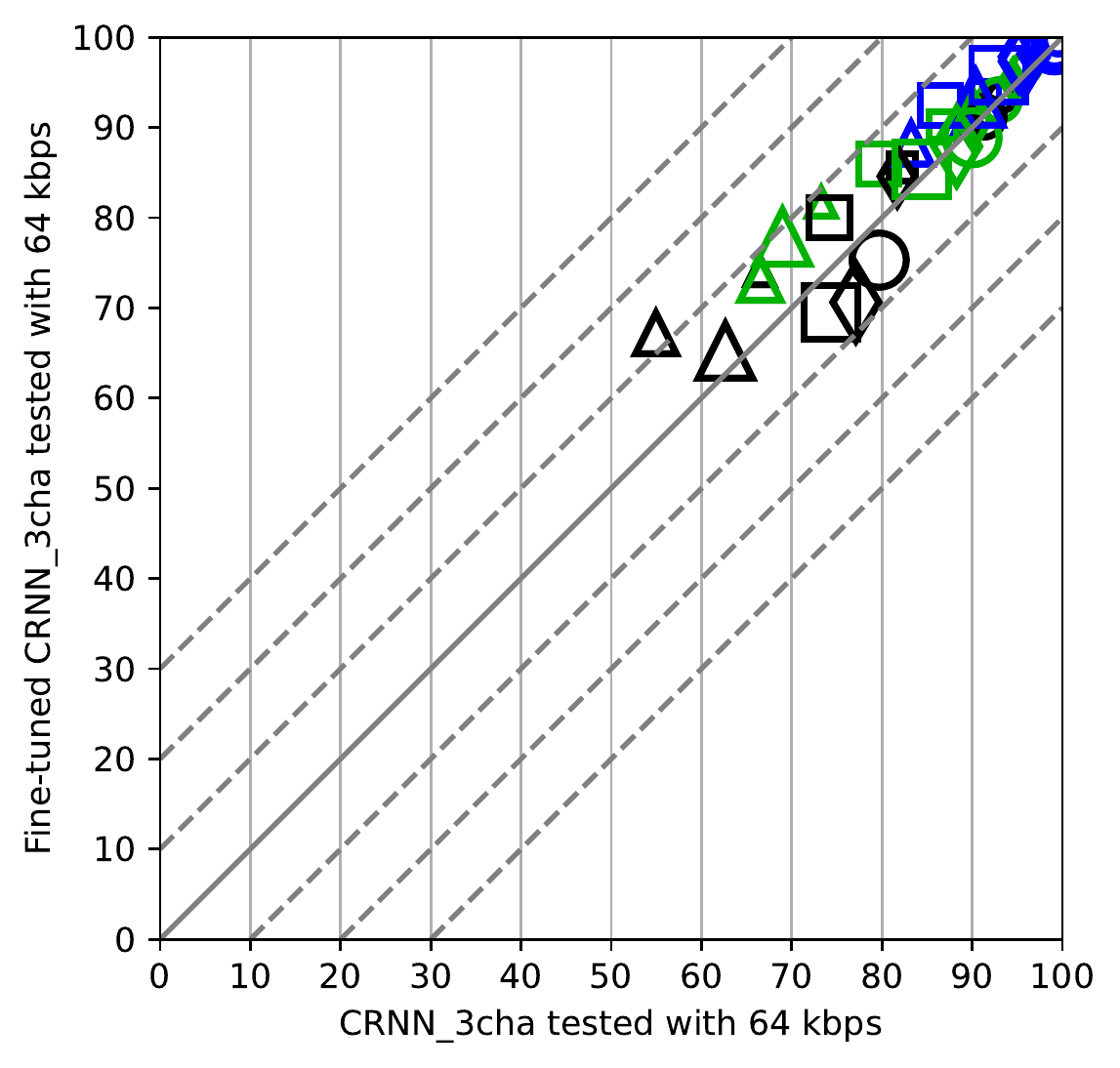}
\label{fig:codec(2)}}
\subfigure{\includegraphics[width=.33\linewidth]{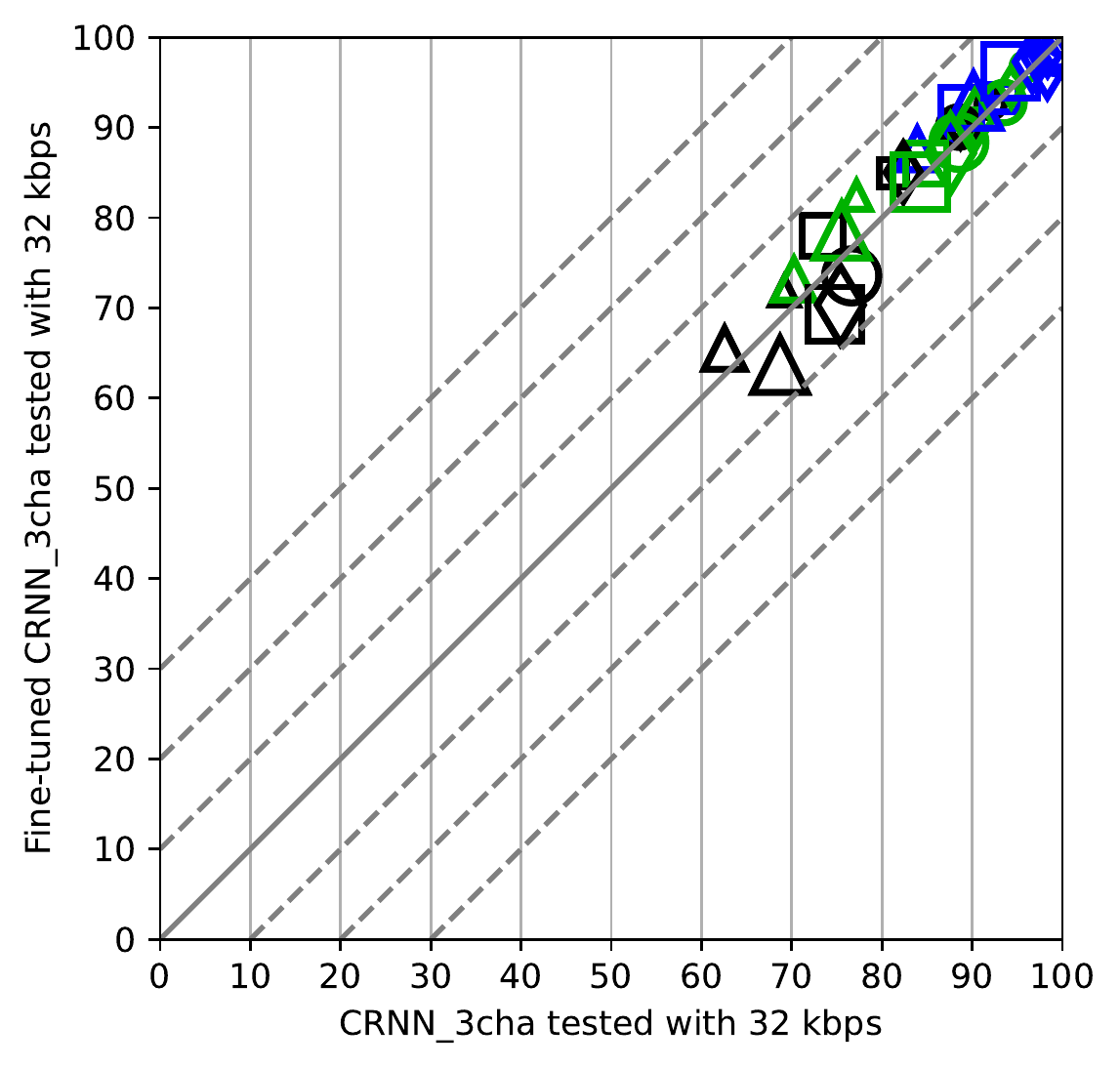}
\label{fig:codec(3)}}
\subfigure{\includegraphics[width=.33\linewidth]{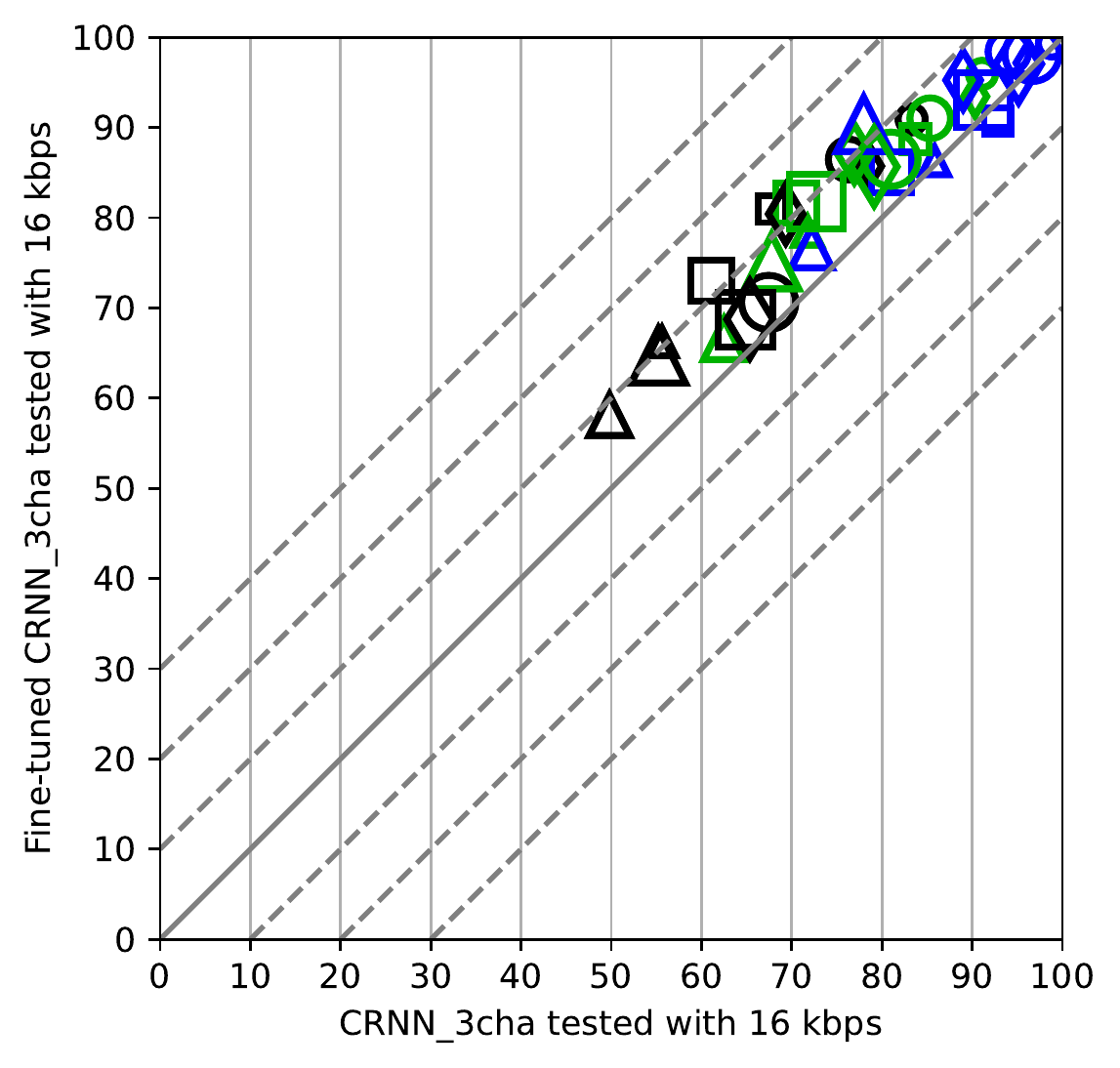}
\label{fig:codec(4)}}
\caption{\label{fig:codec_finetune}
The localisation \textbf{accuracies(\%) } of CRNN baseline and CRNN\_3cha before and after fine-tuning are indicated in the figure. The x-axis in all subplots is model before fine-tuning and y-axis is model after fine-tuning. Note that the models before fine-tunning is trained with 32 kbps encoding and the fine-tuning includes the bitrates of 16, 32, and 64 kbps.}

\end{figure*}
\subsection{BRIR}

BRIR provides the binaural representation of the sound which not only contains the room characteristics for specific listener and source configurations, but also the listener-specific features.
It can be interpreted as the combination of the temporal dynamic of an echoic environment - RIR, with the hearing-device and head-related transfer functions (HD-HRTFs) of the specific listener. To obtain a sufficiently large BRIR dataset for training, instead of recording BRIRs for each combination of parameters, room acoustic simulation methods are utilized to generate synthetic, psycho-acoustically convincing BRIRs that generalise well to real-life conditions \cite{KuttruffRoomAcoustics}. We extend the Pyroomacoustics library \cite{DBLP:journals/corr/abs-1710-04196} to integrate the RIR simulator with different HRTF sets described in Spatially Oriented Format for Acoustics (SOFA)\cite{majdak2013spatially}. The implementation is based on the existing image-source model \cite{ISM1979} by spatial discretisation of the image-source locations to the nearest provided HRTF for all elements in the microphone array. 

To make the localisation model generalise to different rooms, in total 15 rooms of different dimensions and T60s are considered. This is summarised in \tabref{tab:data_generation}.

\subsection{Dataset generation} 

For the training and validation set, as illustrated in \tabref{tab:data_generation}, dry speech sources from different speakers included in the TIMIT \cite{garofolo1993timit} and PTDB-TUG \cite{pirker2011pitch} databases are adopted. 7438 utterances are available for training, and 2280 for validation, with no overlap in terms of the speakers. Training data is generated in the same manner as in~\cite{9381644}. We generate samples in \qty{2}{\second} segments, and in each segment either zero, one, or two sources can be concurrently active\cite{9381644}. A Markov model with two states ({\em active} and {\em inactive}) determines when a source is active. A transition between states occurs on average every \qty{1.5}{\second}. Each time a source becomes active, a random position is chosen from the available BRIRs. The location of each speaker is constant until the source becomes inactive (silent). Changing the source activity is important because we want the recurrent layer to learn the time-variant nature of source activity in real-world settings. Note that we do not simulate gradually moving sources, only sources 'jumping' from location to location. However, we have shown that with the appropriate adaptation to the training data, the network is also able to deal with those types of situations \cite{IMPROVED2023Bohlender}.
The training set comprises BRIRs of 10 different rooms with different dimensions and reverberation times (RT60). 
Spatially diffuse but temporally uncorrelated noise is added with SNRs uniformly ranging from \qty{0}{\decibel} till \qty{30}{\decibel}. We utilize the model of \cite{habets2007generating}. 

The same way of data generation is used to obtain the test-set, but with speech signals from an entirely different database, different BRIRs, and noise conditions. The detailed setting can also be found in \tabref{tab:data_generation}


\section{Results \& Discussion}\label{sec:discussion}

To evaluate the effect of the codec on DoA estimation, the test set is processed with/without LC3plus codec. Three models are tested in this section. 
\subsection{CRNN Baseline}

The performance of CRNN in the baseline condition (no codec applied to any channel, full-bandwidth data exchange) is shown in Fig.\ref{fig:CNN/LSTM} and contrasted, simultaneously, with the two data-exchange models previously described, employing the LC3plus codec. The legend shows the conditions of different SNRs (marker shapes), source-to-array distances (marker sizes), and the number of sources (marker color). Each marker represents one of the corresponding conditions. In both subplots, the x-axis represents the localisation accuracies of the baseline (without encoding). The generally high accuracy scores on this axis confirms our hypothesis that CRNN, in general, works well also for binaural multi-source localisation and, thus, forms a good baseline. Further, the effect of the SNR on the localisation accuracy is significant: the accuracies in noiseless conditions are generally higher than at lower SNRs. 

The y-axis in the left and right subplots represents the test condition with encoding of 3 channels and encoding of 6 channels respectively. All the data points are located under the main diagonal, which indicates that LC3plus degrades the localisation performance. This is not surprising since the encoded signals cause a mismatch between the training and test setups.

\subsection{CRNN trained with codec in loop} \label{sec:codec_discussion}

Two other models with encoding are training for the comparison: 1) CRNN trained with encoding of 3 channels (CRNN\_3cha), and 2) CRNN trained with encoding of 6 channels (CRNN\_6cha). We compare both models with the baseline that was tested without codec to analyse the influence of the LC3plus codec.

The evaluation results are in Fig.\ref{fig:codec_results}. The x-axis in both subplots is the baseline tested without codec. In the left subplot, the y-axis is the CRNN\_3cha tested with encoding of 3 channels. In the right subplot, the y-axis is the CRNN\_6cha tested with encoding of 6 channels. It shows the same tendency in both subplots that the localisation accuracies remain approximately the same in all tested conditions, which proves that DoA information is fully recovered by training with the encoded signal. Surprisingly, in noisy conditions of both subplots, especially SNR = 5 dB, the performance is even {\em improved} by training with the encoded signal. When there is no noise, the performance slightly degrades. Since LC3plus brings distortion in signal, these results suggest that training with such encoding may improve the robustness to other kinds of distortions such as noise. 

There is no significant difference observed by comparing encoding of 3 channels and 6 channels. The DoA information is preserved in both cases.

\subsection{Fine-tuning of different codec bitrates}

To further improve the model robustness of different encoding bitrates, CRNN\_3cha and CRNN\_6cha, which is exclusively trained for 32 kbps, are fine-tuned with the training set encoded with different bitrates: 16, 32, and 64 kbps. Considering the page limit, we take the results of CRNN\_3cha as an example in Fig.\ref{fig:codec_finetune}. The x-axis in all subplots is CRNN\_3cha and the y-axis are fine-tuned CRNN\_3cha, which are tested with the encoding of 64 kbps, 32 kbps, and 16 kbps respectively. 

It is observed that, CRNN\_3cha shows a promising generalization on 16 kbps and 64 kbps signals, which means though training with specific bitrate, it can still generalize between different bitrates. With further fine-tuning, most of the markers in Fig.\ref{fig:codec_finetune} are located above the main diagonal. Also, in the higher SNR scenario, for example 5dB, the improvement is more obvious with fine-tuning. It is proved that fine-tuning among different bitrates helps to improve the localisation accuracy, since it brings more variety of distortion in training. For the specifically trained bitrate 32 kbps (in the middel subplot), there is no significant performance degradation observed though the model is generalized to different bitrates. Especially when testing with encoding of 16 kbps (in the right subplot), fine-tuning improved the accuracies in all tested conditions.

\section{Conclusion}\label{sec:conclusion}
A binaural sound source localisation method CRNN is presented in this paper. The BRIRs of the utilised BTE array are generated with HD-HRTFs recorded in different rooms, whereby the head effect is considered during DoA estimation. The CRNN shows high accuracies with concurrent sources in noisy conditions, which is considered as the baseline to explore the influence of LC3plus codec. Two kinds of encoding methods corresponding to the practical application are considered: encoding of 3 channels and encoding of 6 channels. Based on evaluation results, the conclusion is made that codec processing indeed affects the DOA information, but can still be perfectly recovered by CRNN by training with encoded signals, even bringing robustness in noisy conditions. It is also proved that the models trained with encoding data are robust to different bitrates, which brings a wide possibility of utilizing encoding for binaural localisation. Further work includes using feature-level data exchange instead of audio exchange, based on the model presented in \cite{9657509}, which might lower the data rate even further.

\bibliography{FA2023_template}

\end{document}